&LateX

\documentclass{ws-ijmpa}
\usepackage[super,compress]{cite}
\usepackage{graphicx}
\begin{document}
\markboth{C.T. Potter}{NMSSM Light Decoupled Higgs Singlet at CEPC}

%
\catchline{}{}{}{}{}
%

\title{NMSSM Light Decoupled Higgs Singlet at CEPC}

\author{C.T. Potter}

\address{Physics Department, University of Oregon \\
1274 University of Oregon, Eugene, Oregon 97403-1274, USA}

\maketitle


\begin{abstract}
We describe the phenomenology of light singlet Higgs bosons in the Next-to-Minimal Supersymmetry Model (NMSSM) which are mostly decoupled from the rest of Supersymmetry. Noting that the Large Hadron Collider has not excluded this scenario, we describe previous searches for light Higgs bosons at the Large Electron Positron collider and evaluate the sensitivity to neutralino production and decay to light singlet Higgs bosons at the proposed $\sqrt{s}=250$~GeV Circular Electron Positron Collider.
\end{abstract}

\ccode{PACS numbers: 12.60.Jv, 12.60.Fr, 14.80.Da, 14.80.Nb}

\section{Introduction}

The discovery at the Large Hadron Collider (LHC) of the $h_{125}$ particle \cite{Aad:2012tfa,Chatrchyan:2012ufa} which plays the role of the Standard Model (SM) Higgs boson placed the physics community on the threshold of a much greater potential discovery. 

Supersymmetry (SUSY) posits bosonic partners of every SM fermion and fermionic partners for every SM boson  in order to solve the hierarchy problem (see Ref. \citen{Martin:1997ns} for a review). If discovered, SUSY may also solve residual experimental problems in the SM like the existence of dark matter, the strong CP problem and the anomalous muon magnetic moment.

In addition to the new fermion and vector boson partners, the minimal SUSY model (MSSM) requires two Higgs doublets $H_{u}$ and $H_{d}$, resulting in five physical Higgs states: two CP even neutral states ($h,H$), one CP-odd neutral state ($A$) and two charged states ($H^{\pm}$) \cite{Carena:2002es}. Fermionic partners of the scalar Higgs states complete the MSSM particle landscape. 

The measured properties of the $h_{125}$  \cite{Khachatryan:2016vau} identify it, within experimental error, as the SM Higgs boson and place the other MSSM Higgs bosons well into the decoupling limit $m_{A} \gg m_{Z}$, close to the TeV scale or above. If the MSSM is realized in nature, the recently discovered $h_{125}$ is the $h$ of the MSSM and the $H,A$ and $H^{\pm}$ are undiscovered.

Outside the Higgs sector of the MSSM, LHC searches for SUSY partners of the top quark and the gluon in standard decay channels have imposed severe constraints below the TeV scale (see, for example, Ref. \citen{Aad:2015iea}). This in turn constrains the principle of naturalness, which implies SUSY top and gluon partners below the TeV scale \cite{Papucci:2011wy}. The MSSM also suffers from some theoretical problems, most notably the  inability to account for why, as naturalness requires, the Higgsino mass term $\mu$ is near the electroweak scale \cite{Martin:1997ns}.

The next-to-minimal SUSY model (NMSSM) solves the $\mu$-term problem by introducing an additional Higgs singlet $S$ to the MSSM \cite{Ellwanger:2009dp,Maniatis:2009re}. The $\mu$-term is generated at the electroweak scale by the vacuum expectation value of $S$. In the NMSSM there are seven physical Higgs states: three CP even neutral states ($h_1,h_2,h_3$), two CP-odd neutral states ($a_1,a_2$) and two charged states ($H^{\pm}$). The NMSSM can avoid the experimental constraints on the MSSM by introducing new decays channels with phenomenologies which have not been considered by the LHC experiments.

The NMSSM with a light, mostly decoupled singlet, is one notable example. When the doublet-singlet interaction term  is negligible, the singlet decouples from the rest of SUSY. The $h_{125}$ can be either the $h_1$ or $h_2$ of the NMSSM. If the singlet is light enough, the $h_{125}$ is the $h_2$ of the NMSSM. In addition to two singlet Higgs bosons $a_1$ and $h_1$ in the NMSSM there is an additional neutralino, the singlino. All may be lighter than 125 GeV.

In the following sections, we evaluate the sensitivity of a potential new $e^+ e^-$ collider to light decoupled NMSSM singlet Higgs bosons $a_1$ and $h_1$. Direct production of singlet pairs $e^+ e^- \rightarrow a_1 h_1$ and singlet Higgstrahlung $e^+ e^- \rightarrow Z h_1$ are challenging when the singlet is mostly decoupled, but indirect production from light neutralino decay is a promising channel. Neutralino pair production proceeds through $t$-channel selectron exchange and $s$-channel $Z/\gamma$ exchange. 
After demonstrating that light neutralinos and selectrons have not been excluded by the LHC, we consider the sensitivity to this scenario at the proposed Circular Electron Positron Collider (CEPC) \cite{CEPC-SPPCStudyGroup:2015esa} at $\sqrt{s}=250$~GeV with 10ab$^{-1}$ integrated luminosity.

\section{Phenomenology}

At tree level, the NMSSM Higgs sector is determined by six free parameters \cite{Ellwanger:2009dp}. The parameters $\lambda$ and $\kappa$ appear in the additional terms the NMSSM adds to the MSSM Lagrangian, $\lambda \hat{S} \hat{H}_{d} \hat{H}_{u}+\kappa \hat{S}^3$, the singlet-doublet and singlet self-interaction terms. As noted, the parameter $\mu = \lambda \langle S \rangle$. The parameters $m_{P}$ and $m_{A}$ are the pseudoscalar mass scales. Finally, as for the MSSM, $\tan \beta$ is the ratio of doublet vacuum expectation values.

When $\kappa=0$, the NMSSM exhibits a Peccei-Quinn (PQ) symmetry, and when it is slightly nonzero it exhibits a slightly broken PQ symmetry; when $\lambda\approx 0$ and $\kappa \approx0$ the NMSSM is known as the \emph{effective} MSSM since the MSSM Lagrangian is recovered in the limit $\lambda, \kappa \rightarrow 0$ \cite{Ellwanger:2009dp}. Nevertheless, the phenomeology of an effective MSSM can be strikingly different from the MSSM if the mass hierarchy and $R$-parity conservation forces decays from the NMSSM sector to the SM and MSSM sectors, and \emph{vice versa}.

For $\lambda=0$, the singlet sector is completely decoupled from the MSSM and SM sectors. If we consider $a_1$ to be the lightest singlet sector particle, then it is stable. When $h_1$ decay to singlinos is kinematically disallowed, the only allowed decay is $h_1 \rightarrow a_1 a_1$. Moreover the heavier neutralinos cannot decay to the singlino. However, if a slightly nonzero $\lambda$ is allowed, the $a_1$ and $h_1$ may decay to SM pairs and the MSSM neutralinos may decay to the singlet sector. 

For a concrete example of this scenario, we consider the NMSSM benchmark $h_{60}$ \cite{refId0}, also referred to as BP3 in Ref. \citen{deFlorian:2016spz}. The point was selected from points surviving a large parameter space scan using NMSSMTools 4.4.0 \cite{Ellwanger:2004xm,Ellwanger:2005dv,Belanger:2005kh,Ellwanger:2006rn,Das:2011dg,Muhlleitner:2003vg}, which imposed a large set of experimental constraints, including the $h_{125}$ signal strengths measured at the LHC. 

\begin{table}[h]
\tbl{Masses, dominant decays and branching ratios, and Higgs doublet and singlet components of $a_1,h_1,\chi_1,\chi_2,\chi_3$ in the NMSSM benchmark $h_{60}$.  Note that $a_1,h_1$ and $\chi_1$ are mostly singlet.}
{\begin{tabular}{@{}ccccccc@{}} \toprule
Particle   & Mass [GeV] & Decay & BR [\%] & $H_u$ & $H_{d}$ & $S$  \\ \hline
$a_1$ & 10. & $\tau^+ \tau^-$ & 81 & -0.01 & 0.00 & 1.0  \\
$h_1$ & 56. & $a_1 a_1$ & 72 & -0.01 & -0.13 & 0.99  \\ 
$\chi_{1}$ & 58. & -  & - & 0.15 & -0.05 & 0.98 \\
$\chi_{2}$ & 70. & $\chi_{1} a_{1}$ & 75 & 0.88 & -0.15 & -0.19    \\
$\chi_{3}$ & 122.  & $\chi_{1} h_{1}$ & 80 & 0.40 & 0.68 & 0.05  \\ \botrule
\end{tabular}
\label{tab:h60}}
\end{table}

This point features $\lambda=0.035$ and $\kappa=0.0061$, clearly defining an effective MSSM. The mass scales are $m_{P}=10$~GeV, $m_{A}=1070$~GeV and $\mu=167$~GeV. Finally, $\tan \beta=15$. See Table~\ref{tab:h60} for the masses, main decays and Higgs components of the $h_{60}$ singlet sector $a_1,h_1,\chi_1$ and MSSM sector $\chi_2,\chi_3$. In addition to these light neutrals, there is a light chargino in $h_{60}$ with mass $m_{\chi_1^+}=116$~GeV, which decays dominantly via $\chi^{+}_{1} \rightarrow \chi_2 W^{\star +}$, and a light $m_{\tilde{e}}=205$~GeV selectron which decays to electroweakinos via $\tilde{e}^{\pm} \rightarrow e^{\pm} \chi, \nu \chi^{\pm}$.

\subsection{LHC}

Electroweakino and selectron production with NMSSM  decays as in $h_{60}$ have not been explicitly sought at the LHC experiments. The Runs 1 and 2 exclusion $r_{max}=\sigma_{max}/\sigma_{h_{60}}$ was evaluated with CheckMATE1 1.2.2 \cite{Drees:2013wra} and CheckMATE2 2.0.3 \cite{Dercks:2016npn}. With all currently validated analyses, $r_{max}<1$. See Table~\ref{tab:checkmate}. Maximum exclusion analyses are documented in Refs. \citen{ATLAS-CONF-2013-035,Chatrchyan:2013mys,Khachatryan:2015lwa,ATLAS-CONF-2016-096,ATLAS-CONF-2013-049}.

\begin{table}[h]
\tbl{Maximum CheckMATE exclusion of all validated analyses for $h_{60}$ electroweakino and selectron production at the LHC.}
{\begin{tabular}{@{}ccccc@{}} \toprule
$r_{max}$ &  Analysis & $\sqrt{s}$ [GeV]   & $\int dt \mathcal{L}$ [fb$^{-1}$] & Reference\\ \hline
\multicolumn{5}{c}{Electroweakino $pp \rightarrow \chi \chi,\chi \chi^{\pm},\chi^{+}\chi^{-} $} \\ \hline
0.6 & atlas\_conf\_2013\_035 & 8 & 20.7 & \citen{ATLAS-CONF-2013-035}\\
0.1 & cms\_1303\_2985 & 8 & 11.7 & \citen{Chatrchyan:2013mys}\\ 
0.2 & atlas\_conf\_2016\_096 & 13 & 13.3& \citen{ATLAS-CONF-2016-096}\\ \hline
\multicolumn{5}{c}{Selectron $pp \rightarrow \tilde{e}^{+} \tilde{e}^{-},\tilde{e}^{+}\tilde{\nu},\tilde{e}^{-}\tilde{\nu}$} \\ \hline
0.3 & atlas\_conf\_2013\_049 & 8 & 20.3 & \citen{ATLAS-CONF-2013-049}\\
0.0 & cms\_1502\_06031 & 8 & 19.5 & \citen{Khachatryan:2015lwa}\\ 
0.4 & atlas\_conf\_2016\_096 & 13 & 13.3 & \citen{ATLAS-CONF-2016-096}\\ \botrule
\end{tabular}
\label{tab:checkmate}}
\end{table}

The $h_{60}$ gluino with $m_{\tilde{g}} =611$~GeV was excluded in Run 1. Nevertheless, it was noted \cite{refId0} that the gaugino mass parameter $M_3$, which determines the gluino mass, can be arbitrarily increased beyond LHC exclusion without substantively changing the rest of the $h_{60}$ phenomenology - though possibly with an adverse impact on the naturalness of the point. We also note here that it can be decreased somewhat below kinematic decay thresholds, making it challenging to observe at the LHC. Similar remarks apply to the stop sector.

\subsection{LEP}

Light Higgs bosons of the MSSM have been sought in $e^+ e^-$ collisions at the Large Electron Positron (LEP) collider. At the $Z$ pole $\sqrt{s}=91$~GeV, LEP1 experiments reported exclusion of $e^+ e^- \rightarrow Ah$ \cite{Alexander:1996ai,Adriani:1992kt}. However this MSSM process must be interpreted as doublet Higgs production $e^+ e^- \rightarrow a_2 h_2$ in the $h_{60}$ benchmark, which is consistent with the LHC results $m_{h_2} \approx 125$~GeV and the MSSM decoupling limit $m_{a_2} \gg m_{Z}$. 

The singlet Higgs pair production process $e^+ e^- \rightarrow a_1 h_1$ proceeds in the $s$-channel with a cross section proportional to the square of the coupling $g_{Za_1 h_1}$, given by  \cite{Franke:1995tc}

\begin{eqnarray}
g_{Za_1 h_1} & = & \frac{g}{2 \cos \theta_W} \left( S_{11}P_{11} - S_{12}P_{12} \right)
\end{eqnarray}

\noindent where $S_{11}$ ($S_{12}$) is the $H_u$ ($H_d$) component of the $h_1$ and $P_{11}$ ($P_{12}$) is the $H_u$ ($H_d$) component of the $a_1$. For an effective MSSM with a singlet mostly decoupled from the doublet, this coupling is small. From Table~\ref{tab:h60} these doublet components yield $g^{2}_{Za_1 h_1}/g^2 \approx 10^{-10}$, making $h_{60}$ singlet pair production at the $Z$ pole very challenging for an $e^+ e^-$ collider.

LEP2 experiments also searched for Higgstrahlung $e^+ e^- \rightarrow Z^{(\star)}h$ up to $\sqrt{s}=208$~GeV with null results  \cite{Barate:2003sz}, but only in SM and MSSM decay channels. However the cross section for the NMSSM process $e^+e^- \rightarrow Z^{(\star)} h_1$ in $h_{60}$ is small since the coupling $g_{ZZh_1}$  is given by \cite{Franke:1995tc}

\begin{eqnarray}
g_{ZZh_1} & = & \frac{igm_{Z}}{\cos \theta_W} \left( \cos \beta S_{11}+ \sin \beta S_{12} \right)
\end{eqnarray}

\noindent where $S_{11}$ and $S_{12}$ are as above. For large $\tan \beta$, $\vert g_{ZZh_1}/g_{ZZh_{SM}}\vert^2 \approx S_{12}^2$, so singlet Higgstrahlung in $h_{60}$ is suppressed relative to the SM Higgstrahlung by $S_{12}^2 \approx 0.02$.

But for somewhat larger doublet component $S_{12}$ singlet Higgstrahlung becomes observable at LEP2 energies. The three neutral scalar Higgs bosons in the NMSSM satisfy the coupling sum rule given by \cite{Ellwanger:2009dp} 

\begin{eqnarray}
 \sum_{i} g^{2}_{ZZh_i} = & g^{2}_{ZZh_{SM}}
\end{eqnarray}

\noindent thereby, when $h_2$ is identified with $h_{125}$, constraining $g_{ZZh_1}$ and $g_{ZZh_3}$  due the SM-like signal strengths for $h_{125} \rightarrow ZZ^{\star}$ measured at the LHC. The combined Run 1 ATLAS and CMS $1\sigma$ lower bound on the coupling modifier $\kappa_{Z}=g_{ZZh_{125}}/g_{ZZh_{SM}}$ is $\kappa_{Z} > 0.90$ \cite{Khachatryan:2016vau}, which imposes the upper bound $\vert S_{12}\vert <0.44$ for large $\tan \beta$.

A recent re-analyis of LEP2 data from ALEPH has noted an excess of order 100 hadronic events with four jets and dijet masses $M_1^{\prime}\approx 90-100$~GeV, $M_{2}^{\prime} \approx 25$~GeV \cite{Kile:2017ccn}. Singlet Higgstrahlung $e^+ e^- \rightarrow  Z h_1$ with $h_1 \rightarrow 2a_1 \rightarrow 4\tau$ may be consistent with the excess topology, depending on details of the jetfinding. In $h_{60}$ the number of $Zh_1$ events produced in the $\sqrt{s},\int dt \mathcal{L}$ profile of the ALEPH dataset \cite{Kile:2017ccn} yields approximately 13 events. An enhancement of $S_{12}$ by a factor three could account for the excess and still satisfy the LHC constraint $\vert S_{12}\vert <0.44$.

\subsection{CEPC}

In addition to direct production, the $a_1$ and $h_1$ may also be produced in electroweakino decays. When doublet dominated NMSSM electroweakino masses are below threshold for decay to SUSY particles outside the singlet sector,  $R$-parity conservation forces decay to singlets and singlinos. 

\begin{figure}[h]
\begin{center}
\includegraphics[width=0.8\textwidth]{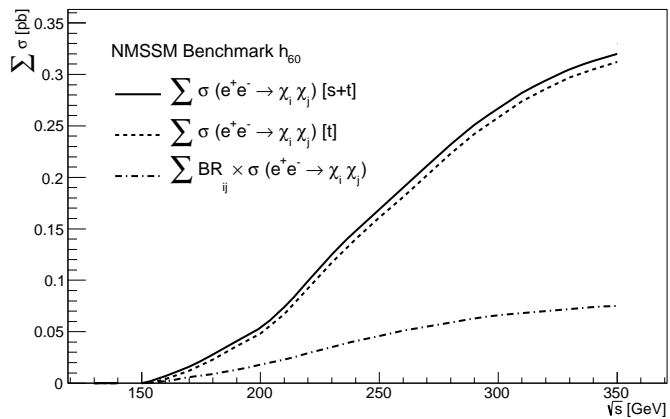}
\caption{The MG5\_aMC@NLO cross section $\sum_{i,j} \sigma(e^{+}e^{-} \rightarrow \chi_i \chi_j)$ ($i=1,2,3$ and $j=2,3$) vs. $\sqrt{s}$ for $h_{60}$. The solid line includes $s$- and $t$-channels, the dashed line includes only the $t$-channel, and the dot-dash line includes both channels including branching ratios to dominant decays.}
\label{fig:xsec}
\end{center}
\end{figure}

Neutralino pair production occurs through $Z/\gamma$ $s$-channel and $t$-channel selectron exchange. See Figure \ref{fig:xsec} for the total $h_{60}$ neutralino pair production cross section versus $\sqrt{s}$ obtained with MG5\_aMC@NLO 2.3.3 \cite{Alwall:2014hca}.   In $h_{60}$, the processes $e^+ e^- \rightarrow \chi_{i} \chi_{j}$ ($i=1,2,3$ and $j=2,3$) at $\sqrt{s}=250$~GeV dominantly produce two $\chi_1$ and between one and four $a_1$ with $BR(a_1 \rightarrow \tau^+ \tau^-) =0.81$.  

If the singlet is too loosely coupled, the $\chi_2$ and $\chi_3$ widths may be small enough that the neutralinos decay outside the effective tracking volume of a detector. In $h_{60}$ decay distances $c\tau \approx 0.1\mu$m (2pm) for the $\chi_2$ ($\chi_3$), well inside the tracking radius. Similarly, $c\tau\approx 10$nm for the $a_1$, ensuring detection of charged tracks from $a_1 \rightarrow \tau^+ \tau^-$.

The process $e^+ e^- \rightarrow  \chi_{1}^{+} \chi_{1}^{-}$ produces two $\chi_1$  and two $a_1$, but also two offshell $W^{\star}$  through the decay $\chi_{1}^{+} \rightarrow \chi_2 W^{\star}$ making it challenging to separate from background. In the following section we focus on neutralino pair production and leave the chargino channel for a dedicated future study.

\section{Analysis}

To demonstrate CEPC sensitivity to the light decoupled Higgs scenario discussed above, we analyze signal and background simulations generated in MG5\_aMC@NLO 2.3.3 assuming $\sqrt{s}=250$~GeV, $\int dt \mathcal{L}=10$ab$^{-1}$ and unpolarized $e^+ e^-$ beams. In principle the results also apply for the International Linear Collider (ILC) \cite{Behnke:2013lya}, though run scenarios for the ILC typically assume polarized beams and lower luminosity \cite{Barklow:2015tja}.

Fast detector simulation is performed by Delphes 3.3.1 \cite{Selvaggi:2014mya,Mertens:2015kba}. We use the DSiD Delphes card \cite{Potter:2016pgp}, which emulates the full simulation performance of the SiD detector \cite{Behnke:2013lya}. SiD was designed as a detector for the ILC, but has also been proposed as a detector for CEPC \cite{Chekanov:2016efe}. 

In the simulated signal sample we generate $2.5 \times 10^6$ $e^+ e^- \rightarrow \chi_{i} \chi_{j}$  events requiring the $h_{60}$ dominant decays for each particle in each event (see Table~\ref{tab:h60}). Using the $h_{60}$ SLHA \cite{Skands:2003cj} file, MG5\_aMC@NLO yields $\sum \sigma_{\chi \chi} \times BR \approx 51$~fb. Unconstrained tau decays are performed by Pythia6 \cite{Sjostrand:2006za} libraries using Tauola \cite{Was:2011tv}. 

For a detailed description of the background samples used here, see Ref. \citen{Potter:2017rlo}. Briefly, these include initial states $e^+ e^-$,$\gamma \gamma$ with difermion and diboson final states $f\bar{f}$, $VV$. They also include $e \gamma$ initial states with final states $eV$,$\nu V$. For each sample, enough events were produced such that their equivalent integrated luminosity is 10ab$^{-1}$ or more. The MG5\_aMC@NLO cross sections are used to normalize background events, except when they conflict with Ref. \citen{Mo:2015mza}, which include radiative effects tailored to CEPC, in which case the latter cross sections are used.

Approximately 46\% of taus decay to single charged particles and one or two neutrinos, $\tau^+ \rightarrow e^+ \nu_{\tau} \nu_{e}, \mu^+ \nu_{\tau} \nu_{\mu}, \pi^+ \nu_{\tau}$, while another 35\% of taus decay to a single charged pion, one neutrino and one or two neutral pions $\tau^+ \rightarrow \pi^+ \pi^0 \nu_{\tau}, \pi^+ 2\pi^0 \nu_{\tau}$. An additional 4\% of taus decay in other so-called \emph{one-prong} decays to one track.

In the simulated signal and background samples we make a straightforward selection for the $h_{60}$ neutralino  pair production processes accessible at $\sqrt{s}=250$~GeV. The selection relies heavily on particle flow techniques and highly granulated calorimetry for distinguishing charged and neutral calorimeter clusters, both strong features of the SiD detector.

We select for one isolated $a_1$ decaying to two one-prong taus and recoiling against either one or two $a_1$ also decaying to one-prong taus, resulting in at most six charged particles. We allow some inefficiency in signal track reconstruction, but require at least three charged tracks to eliminate background. We therefore select for $3 \leq N_{trk} \leq 6$ with $p_{T}^{trk}>0.4$~GeV. Since two $\chi_1$ will accompany each event, we require very little visible energy in the event, $E_{vis}<0.2 \sqrt{s}$. To suppress background from two-fermion events like $e^+ e^- \rightarrow \tau^+ \tau^-$, we require the event sphericity $Sph>0.1$. 

For the isolated $a_1$ we require at least one track pair with net charge zero separated by at least $10^{\circ}$ and at most $90^{\circ}$, with zero additional energy from neutral hadrons and tracks in the hemisphere defined by the track pair three vector. The pair with minimal angular separation is chosen for the isolated $a_1$ candidate. Additional photons in the $a_1$ candidate hemisphere are allowed to accommodate decay $\pi^0 \rightarrow \gamma \gamma$ of neutral pions produced in the tau decay. Finally the total invariant masses in both the $a_1$ candidate (same) hemisphere and the opposite hemisphere are required to satisfy $m_{sh},m_{oh}>0.5$~GeV.

\begin{table}[h]
\tbl{Cumulative event yields for the signal selection described in the text assuming $\int dt \mathcal{L}=$10ab$^{-1}$. See the text for track pair selection details.}
{\begin{tabular}{@{}cccc@{}} \toprule
Requirement & Signals S & Background B & $S/\sqrt{B}$ \\  \hline 
None & 510 & $6.61\times 10^5$ & 0.627 \\
$3 \leq N_{trk} \leq 6$ & 371 & $1.56\times 10^4$ & 2.96 \\
$E_{vis}<0.2 \sqrt{s}$ & 362 & $2.48\times 10^3$ & 7.27 \\
$Sph>0.1$  & 272 & $1.12\times 10^3$ & 8.12 \\
Track Pair & 135 & 17.2 & 32.7 \\
$m_{sh}>0.5$~GeV & 112 & 14.3 & 29.5 \\ 
$m_{os}>0.5$~GeV  & 112 & 14.3 & 29.5 \\ \botrule
\end{tabular}
\label{tab:yields}}
\end{table}

\begin{figure}[h]
\begin{center}
\includegraphics[width=0.49\textwidth]{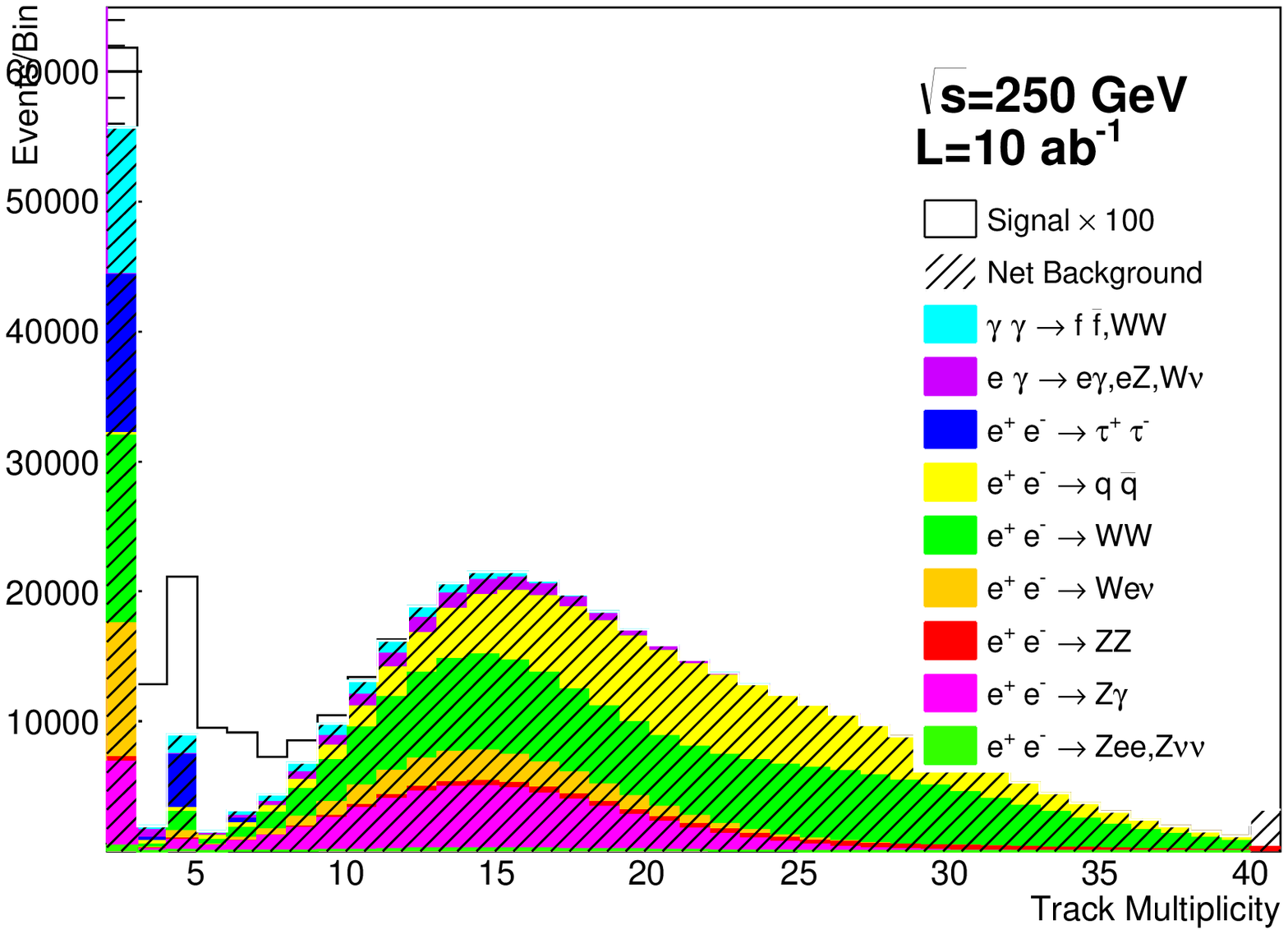}
\includegraphics[width=0.49\textwidth]{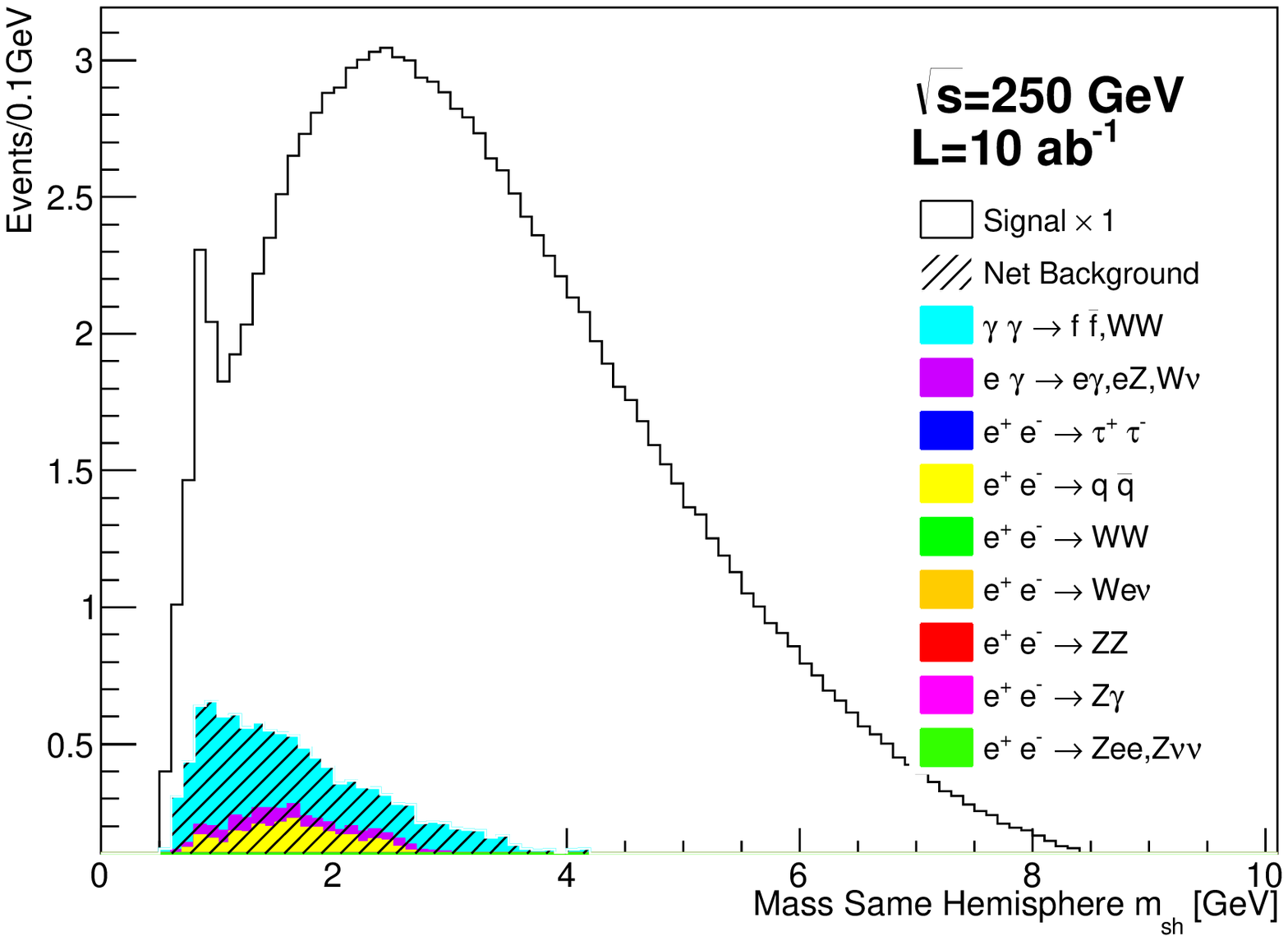}
\caption{The track multiplicity before any signal selection is applied (left) and the mass in the selected $a_1$ candidate hemisphere after full selection (right). In the left plot the signal is enhanced by $\times100$. Signal is white, total background is hatched. The signal feature at low mass is due to $\rho_{770} \rightarrow \pi^+ \pi^-$.}
\label{fig:analysis}
\end{center}
\end{figure}

For the signal and background event yields at each step in the analysis selection, see Table~\ref{tab:yields}. For the track multiplicity distribution before any selection and the invariant mass in the $a_1$ candidate hemisphere after full signal selection, see Figure~\ref{fig:analysis}.  

After full signal selection, 112 signal events remain and 14 background events remain. The dominant background is quark pair events $\gamma \gamma \rightarrow q \bar{q}$ and $e^+ e^- \rightarrow q \bar{q}$ with low track multiplicity and small visible energy from multiple semileptonic meson cascade decays. Background $\gamma \gamma \rightarrow q\bar{q}$ events dominate since they are produced at energies well below the $e^+e^-$ beam $\sqrt{s}$.

It should be noted both that a very pure, if potentially biased, signal sample can be obtained by selecting events with $m_{sh}>4$~GeV, and that the signal selection can be further optimized with, for example, dedicated tau tagging and multivariate techniques.  Finally, signal events with one or more subdominant decay, which have not been considered here, account for an unexplored $2/3$ of the total signal cross section.

\section{Conclusion}

We have described the phenomenology of a light singlet Higgs in the NMSSM which is largely decoupled from the doublet Higgs. Such a scenario remains viable even after Run 2 at the LHC. With $e^+ e^-$ colliders the decoupled light singlet is challenging to observe both in $Z \rightarrow a_1 h_1$ at the LEP1 energy $\sqrt{s}=91$~GeV and singlet Higgstrahlung $e^+ e^- \rightarrow Zh_1$ at LEP2 energies $91 < \sqrt{s} < 208$~GeV.

But with $e^+ e^-$ colliders at $\sqrt{s}=250$~GeV, neutralino pair production with decays to singlet Higgs bosons can proceed with a high enough cross section and low enough background to offer a unique opportunity to study these bosons in a clean event environment. For example, we evaluated the sensitivity of CEPC with $\sqrt{s}=250$~GeV and $\int dt \mathcal{L}=10$ab$^{-1}$ to a benchmark point with a light NMSSM singlet sector $a_1,h_1,\chi_1$ with masses $10,56,58$~GeV. We find after a simple selection a high purity sample can be obtained with more than 100 signal events. The selection described has not been optimized, which is left for a future work.

\section*{Acknowledgments}
The author thanks the Institute for Advanced Study at HKUST for its 2017 Program on High Energy Physics and the Alder Institute for High Energy Physics for financial support.

\bibliography{paper}

\begin{thebibliography}{10}
\expandafter\ifx\csname urlstyle\endcsname\relax
  \providecommand{\doi}[1]{doi:\discretionary{}{}{}#1}\else
  \providecommand{\doi}{doi:\discretionary{}{}{}\begingroup
  \urlstyle{rm}\Url}\fi

\bibitem{Aad:2012tfa}
 ATLAS Collaboration Collaboration (G.~Aad {\em et~al.}), {\em Phys.Lett.} {\bf
  B716}, 1  (2012), \href{http://arxiv.org/abs/1207.7214}{{\ttfamily
  arXiv:1207.7214 [hep-ex]}}, \doi{10.1016/j.physletb.2012.08.020}.

\bibitem{Chatrchyan:2012ufa}
 CMS Collaboration Collaboration (S.~Chatrchyan {\em et~al.}), {\em Phys.Lett.}
  {\bf B716}, 30  (2012), \href{http://arxiv.org/abs/1207.7235}{{\ttfamily
  arXiv:1207.7235 [hep-ex]}}, \doi{10.1016/j.physletb.2012.08.021}.

\bibitem{Martin:1997ns}
S.~P. Martin  (1997), \href{http://arxiv.org/abs/hep-ph/9709356}{{\ttfamily
  arXiv:hep-ph/9709356 [hep-ph]}}.

\bibitem{Carena:2002es}
M.~S. Carena and H.~E. Haber, {\em Prog.Part.Nucl.Phys.} {\bf 50}, 63  (2003),
  \href{http://arxiv.org/abs/hep-ph/0208209}{{\ttfamily arXiv:hep-ph/0208209
  [hep-ph]}}, \doi{10.1016/S0146-6410(02)00177-1}.

\bibitem{Khachatryan:2016vau}
 ATLAS, CMS Collaboration (G.~Aad {\em et~al.}), {\em JHEP} {\bf 08},   045
  (2016), \href{http://arxiv.org/abs/1606.02266}{{\ttfamily arXiv:1606.02266
  [hep-ex]}}, \doi{10.1007/JHEP08(2016)045}.

\bibitem{Aad:2015iea}
 ATLAS Collaboration (G.~Aad {\em et~al.})  (2015),
  \href{http://arxiv.org/abs/1507.05525}{{\ttfamily arXiv:1507.05525
  [hep-ex]}}.

\bibitem{Papucci:2011wy}
M.~Papucci, J.~T. Ruderman and A.~Weiler, {\em JHEP} {\bf 1209},   035  (2012),
  \href{http://arxiv.org/abs/1110.6926}{{\ttfamily arXiv:1110.6926 [hep-ph]}},
  \doi{10.1007/JHEP09(2012)035}.

\bibitem{Ellwanger:2009dp}
U.~Ellwanger, C.~Hugonie and A.~M. Teixeira, {\em Phys.Rept.} {\bf 496}, 1
  (2010), \href{http://arxiv.org/abs/0910.1785}{{\ttfamily arXiv:0910.1785
  [hep-ph]}}, \doi{10.1016/j.physrep.2010.07.001}.

\bibitem{Maniatis:2009re}
M.~Maniatis, {\em Int.J.Mod.Phys.} {\bf A25}, 3505  (2010),
  \href{http://arxiv.org/abs/0906.0777}{{\ttfamily arXiv:0906.0777 [hep-ph]}},
  \doi{10.1142/S0217751X10049827}.

\bibitem{CEPC-SPPCStudyGroup:2015esa}
C.-S.~S. Group  (2015).

\bibitem{refId0}
{Potter, C. T.}, {\em Eur. Phys. J. C} {\bf 76},  ~44  (2016),
  \doi{10.1140/epjc/s10052-015-3867-x}.

\bibitem{deFlorian:2016spz}
 LHC Higgs Cross Section Working Group Collaboration (D.~de~Florian {\em
  et~al.})  (2016), \href{http://arxiv.org/abs/1610.07922}{{\ttfamily
  arXiv:1610.07922 [hep-ph]}}, \doi{10.23731/CYRM-2017-002}.

\bibitem{Ellwanger:2004xm}
U.~Ellwanger, J.~F. Gunion and C.~Hugonie, {\em JHEP} {\bf 0502},   066
  (2005), \href{http://arxiv.org/abs/hep-ph/0406215}{{\ttfamily
  arXiv:hep-ph/0406215 [hep-ph]}}, \doi{10.1088/1126-6708/2005/02/066}.

\bibitem{Ellwanger:2005dv}
U.~Ellwanger and C.~Hugonie, {\em Comput.Phys.Commun.} {\bf 175}, 290  (2006),
  \href{http://arxiv.org/abs/hep-ph/0508022}{{\ttfamily arXiv:hep-ph/0508022
  [hep-ph]}}, \doi{10.1016/j.cpc.2006.04.004}.

\bibitem{Belanger:2005kh}
G.~Belanger, F.~Boudjema, C.~Hugonie, A.~Pukhov and A.~Semenov, {\em JCAP} {\bf
  0509},   001  (2005), \href{http://arxiv.org/abs/hep-ph/0505142}{{\ttfamily
  arXiv:hep-ph/0505142 [hep-ph]}}, \doi{10.1088/1475-7516/2005/09/001}.

\bibitem{Ellwanger:2006rn}
U.~Ellwanger and C.~Hugonie, {\em Comput.Phys.Commun.} {\bf 177}, 399  (2007),
  \href{http://arxiv.org/abs/hep-ph/0612134}{{\ttfamily arXiv:hep-ph/0612134
  [hep-ph]}}, \doi{10.1016/j.cpc.2007.05.001}.

\bibitem{Das:2011dg}
D.~Das, U.~Ellwanger and A.~M. Teixeira, {\em Comput.Phys.Commun.} {\bf 183},
  774  (2012), \href{http://arxiv.org/abs/1106.5633}{{\ttfamily arXiv:1106.5633
  [hep-ph]}}, \doi{10.1016/j.cpc.2011.11.021}.

\bibitem{Muhlleitner:2003vg}
M.~Muhlleitner, A.~Djouadi and Y.~Mambrini, {\em Comput.Phys.Commun.} {\bf
  168}, 46  (2005), \href{http://arxiv.org/abs/hep-ph/0311167}{{\ttfamily
  arXiv:hep-ph/0311167 [hep-ph]}}, \doi{10.1016/j.cpc.2005.01.012}.

\bibitem{Drees:2013wra}
M.~Drees, H.~Dreiner, D.~Schmeier, J.~Tattersall and J.~S. Kim, {\em Comput.
  Phys. Commun.} {\bf 187}, 227  (2014),
  \href{http://arxiv.org/abs/1312.2591}{{\ttfamily arXiv:1312.2591 [hep-ph]}},
  \doi{10.1016/j.cpc.2014.10.018}.

\bibitem{Dercks:2016npn}
D.~Dercks, N.~Desai, J.~S. Kim, K.~Rolbiecki, J.~Tattersall and T.~Weber
  (2016), \href{http://arxiv.org/abs/1611.09856}{{\ttfamily arXiv:1611.09856
  [hep-ph]}}.

\bibitem{ATLAS-CONF-2013-035}
ATLAS Collaboration Collaboration, {\em {Search for direct production of
  charginos and neutralinos in events with three leptons and missing transverse
  momentum in 21$\,$fb$^{-1}$ of pp collisions at $\sqrt{s}=8\,$TeV with the
  ATLAS detector}}, Tech. Rep. ATLAS-CONF-2013-035, CERN (Mar 2013).

\bibitem{Chatrchyan:2013mys}
 CMS Collaboration (S.~Chatrchyan {\em et~al.}), {\em Eur. Phys. J.} {\bf C73},
    2568  (2013), \href{http://arxiv.org/abs/1303.2985}{{\ttfamily
  arXiv:1303.2985 [hep-ex]}}, \doi{10.1140/epjc/s10052-013-2568-6}.

\bibitem{Khachatryan:2015lwa}
 CMS Collaboration (V.~Khachatryan {\em et~al.}), {\em JHEP} {\bf 04},   124
  (2015), \href{http://arxiv.org/abs/1502.06031}{{\ttfamily arXiv:1502.06031
  [hep-ex]}}, \doi{10.1007/JHEP04(2015)124}.

\bibitem{ATLAS-CONF-2016-096}
ATLAS Collaboration Collaboration, {\em {Search for supersymmetry with two and
  three leptons and missing transverse momentum in the final state at
  $\sqrt{s}=13$ TeV with the ATLAS detector}}, Tech. Rep. ATLAS-CONF-2016-096,
  CERN (Sep 2016).

\bibitem{ATLAS-CONF-2013-049}
{\em {Search for direct-slepton and direct-chargino production in final states
  with two opposite-sign leptons, missing transverse momentum and no jets in
  20/fb of pp collisions at sqrt(s) = 8 TeV with the ATLAS detector}}, Tech.
  Rep. ATLAS-CONF-2013-049, CERN (May 2013).

\bibitem{Alexander:1996ai}
 OPAL Collaboration (G.~Alexander {\em et~al.}), {\em Z. Phys.} {\bf C73}, 189
  (1997), \doi{10.1007/s002880050309}.

\bibitem{Adriani:1992kt}
 L3 Collaboration (O.~Adriani {\em et~al.}), {\em Z. Phys.} {\bf C57}, 355
  (1993), \doi{10.1007/BF01474331}.

\bibitem{Franke:1995tc}
F.~Franke and H.~Fraas, {\em Int. J. Mod. Phys.} {\bf A12}, 479  (1997),
  \href{http://arxiv.org/abs/hep-ph/9512366}{{\ttfamily arXiv:hep-ph/9512366
  [hep-ph]}}, \doi{10.1142/S0217751X97000529}.

\bibitem{Barate:2003sz}
 OPAL, DELPHI, LEP Working Group for Higgs boson searches, ALEPH, L3
  Collaboration (R.~Barate {\em et~al.}), {\em Phys. Lett.} {\bf B565}, 61
  (2003), \href{http://arxiv.org/abs/hep-ex/0306033}{{\ttfamily
  arXiv:hep-ex/0306033 [hep-ex]}}, \doi{10.1016/S0370-2693(03)00614-2}.

\bibitem{Kile:2017ccn}
J.~Kile and J.~von Wimmersperg-Toeller  (2017),
  \href{http://arxiv.org/abs/1706.02255}{{\ttfamily arXiv:1706.02255
  [hep-ex]}}.

\bibitem{Alwall:2014hca}
J.~Alwall, R.~Frederix, S.~Frixione, V.~Hirschi, F.~Maltoni, O.~Mattelaer,
  H.~S. Shao, T.~Stelzer, P.~Torrielli and M.~Zaro, {\em JHEP} {\bf 07},   079
  (2014), \href{http://arxiv.org/abs/1405.0301}{{\ttfamily arXiv:1405.0301
  [hep-ph]}}, \doi{10.1007/JHEP07(2014)079}.

\bibitem{Behnke:2013lya}
H.~Abramowicz {\em et~al.}  (2013),
  \href{http://arxiv.org/abs/1306.6329}{{\ttfamily arXiv:1306.6329
  [physics.ins-det]}}.

\bibitem{Barklow:2015tja}
T.~Barklow, J.~Brau, K.~Fujii, J.~Gao, J.~List, N.~Walker and K.~Yokoya
  (2015), \href{http://arxiv.org/abs/1506.07830}{{\ttfamily arXiv:1506.07830
  [hep-ex]}}.

\bibitem{Selvaggi:2014mya}
M.~Selvaggi, {\em J. Phys. Conf. Ser.} {\bf 523},   012033  (2014),
  \doi{10.1088/1742-6596/523/1/012033}.

\bibitem{Mertens:2015kba}
A.~Mertens, {\em J. Phys. Conf. Ser.} {\bf 608},   012045  (2015),
  \doi{10.1088/1742-6596/608/1/012045}.

\bibitem{Potter:2016pgp}
C.~T. Potter, { {DSiD: a Delphes Detector for ILC Physics Studies}}, in {\em
  {Proceedings, International Workshop on Future Linear Colliders (LCWS15):
  Whistler, B.C., Canada, November 02-06, 2015}\/},  (2016).
\newblock \href{http://arxiv.org/abs/1602.07748}{{\ttfamily arXiv:1602.07748
  [hep-ph]}}.

\bibitem{Chekanov:2016efe}
S.~V. Chekanov and M.~Demarteau, {\em Int. J. Mod. Phys.} {\bf A31},   1644021
  (2016), \href{http://arxiv.org/abs/1604.01994}{{\ttfamily arXiv:1604.01994
  [physics.ins-det]}}, \doi{10.1142/S0217751X16440218}.

\bibitem{Skands:2003cj}
P.~Z. Skands {\em et~al.}, {\em JHEP} {\bf 07},   036  (2004),
  \href{http://arxiv.org/abs/hep-ph/0311123}{{\ttfamily arXiv:hep-ph/0311123
  [hep-ph]}}, \doi{10.1088/1126-6708/2004/07/036}.

\bibitem{Sjostrand:2006za}
T.~Sjostrand, S.~Mrenna and P.~Z. Skands, {\em JHEP} {\bf 0605},   026  (2006),
  \href{http://arxiv.org/abs/hep-ph/0603175}{{\ttfamily arXiv:hep-ph/0603175
  [hep-ph]}}, \doi{10.1088/1126-6708/2006/05/026}.

\bibitem{Was:2011tv}
Z.~Was, {\em Nucl. Phys. Proc. Suppl.} {\bf 218}, 249  (2011),
  \href{http://arxiv.org/abs/1101.1652}{{\ttfamily arXiv:1101.1652 [hep-ph]}},
  \doi{10.1016/j.nuclphysbps.2011.06.040}.

\bibitem{Potter:2017rlo}
C.~T. Potter  (2017), \href{http://arxiv.org/abs/1702.04827}{{\ttfamily
  arXiv:1702.04827 [hep-ph]}}.

\bibitem{Mo:2015mza}
X.~Mo, G.~Li, M.-Q. Ruan and X.-C. Lou, {\em Chin. Phys.} {\bf C40},   033001
  (2016), \href{http://arxiv.org/abs/1505.01008}{{\ttfamily arXiv:1505.01008
  [hep-ex]}}, \doi{10.1088/1674-1137/40/3/033001}.

\end{thebibliography}

\end{document}